\documentclass[fleqn,10pt]{wlscirep}
\usepackage[utf8]{inputenc}
\usepackage[T1]{fontenc}
\usepackage[version=4]{mhchem}
\usepackage{bm}
\usepackage{graphicx}
	
\title{Smooth, homogeneous, high-purity Nb$_3$Sn superconducting RF resonant cavity by seed-free electrochemical synthesis}

\author[1,2,$\dagger$,$\star$]{Zeming Sun}
\author[3,$\dagger$]{Zhaslan Baraissov}
\author[1,2]{Ryan D. Porter}
\author[1,2]{Liana Shpani}
\author[3]{Yu-Tsun Shao}
\author[1,2]{Thomas Oseroff}
\author[4]{Michael O. Thompson}
\author[3]{David A. Muller}
\author[1,2]{Matthias U. Liepe}
\affil[1]{Cornell Laboratory for Accelerator-Based Sciences and Education, Cornell University, Ithaca, NY, 14853, United States of America}
\affil[2]{Department of Physics, Cornell University, Ithaca, NY, 14853, United States of America}
\affil[3]{School of Applied and Engineering Physics, Cornell University, Ithaca, NY, 14853, United States of America}
\affil[4]{Department of Materials Science and Engineering, Cornell University, Ithaca, NY, 14853, United States of America}
\affil[$\dagger$]{contributed equally}
\affil[$\star$]{e-mail: zs253@cornell.edu}

\begin{abstract}

Workbench-size particle accelerators, enabled by Nb$_3$Sn-based superconducting radio-frequency (SRF) cavities, hold the potential of driving scientific discovery by offering a widely accessible and affordable source of high-energy electrons and X-rays. Thin-film Nb$_3$Sn RF superconductors with high quality factors, high operation temperatures, and high-field potentials are critical for these devices. However, surface roughness, non-stoichiometry, and impurities in Nb$_3$Sn deposited by conventional Sn-vapor diffusion prevent them from reaching their theoretical capabilities. Here we demonstrate a seed-free electrochemical synthesis that pushes the limit of chemical and physical properties in Nb$_3$Sn. Utilization of electrochemical Sn pre-deposits reduces the roughness of converted Nb$_3$Sn by five times compared to typical vapor-diffused Nb$_3$Sn. Quantitative mappings using chemical and atomic probes confirm improved stoichiometry and minimized impurity concentrations in electrochemically synthesized Nb$_3$Sn. We have successfully applied this Nb$_3$Sn to the large-scale 1.3\,GHz SRF cavity and demonstrated ultra-low BCS surface resistances at multiple operation temperatures, notably lower than vapor-diffused cavities. Our smooth, homogeneous, high-purity Nb$_3$Sn provides the route toward high efficiency and high fields for SRF applications under helium-free cryogenic operations.

\end{abstract}
\begin{document}

\flushbottom
\maketitle

\thispagestyle{empty}

\section{Introduction}

RF (or microwave) superconductors \cite{SunRef72,SunRef85,SunRef20} find use in a wide range of modern technologies, including particle accelerator components (\textit {e.g.}, superconducting radio-frequency (SRF) resonant cavities and photocathodes using SRF guns) \cite{SunRef20,SunRef103,SunRef100,SunRef83,SunRef90,SunRef21,SunRef22,SunRef92}, superconducting quantum circuits \cite{SunRef17,SunRef28} and superconducting parametric readout amplifiers \cite{SunRef80}, superconducting quantum photonics \cite{SunRef82}, and ultra-sensitive detectors and filters (\textit {e.g.}, superconducting transition-edge sensors \cite{SunRef81} and kinetic inductance detectors \cite{SunRef86}). An immediate family of these applications spans equally a broad range of disciplines in materials science (\textit {e.g.}, structural characterization, atomic analysis, and thermodynamic/kinetic studies) \cite{SunRef83,SunRef81,SunRef21,SunRef22}, photon science (synchrotrons and free-electron lasers \cite{SunRef21,SunRef22,SunRef103,SunRef83,SunRef100}) and ultrafast MeV electron microscopes \cite{SunRef92}, physics (high-energy physics \cite{SunRef90}, particle and nuclear physics \cite{SunRef88,SunRef81}, isotope technology \cite{SunRef91}, quantum technology \cite{SunRef17,SunRef28,SunRef26}, and astrophysics and dark matter detection \cite{SunRef86,SunRef81,SunRef23}), chemistry and biology (\textit {e.g.}, molecular vibration, chemical bonding, elemental analysis, reaction and kinetics, live-cell imaging, and biological material analysis) \cite{SunRef81,SunRef83,SunRef21,SunRef22,SunRef103}, and medical and biopharmaceutical applications (\textit {e.g.}, cancer therapy) \cite{SunRef87}. 

Currently, applications of modern accelerators, such as high-energy electrons and X-rays, are limited to kilometer-scale large facilities, with around 50 sites worldwide, including SLAC LCLS, SHINE, European XFEL, among others. The transition of accelerators from low-gradient normal-conducting RF to high-gradient superconducting RF has made bulk Nb the dominant cavity technology. Recently, efforts have been focused on scaling down large accelerator facilities and exploring lab-scale dimensions \cite{SunRef34,SunRef106,SunRef107,SunRef108}. To achieve a compact, turn-key, and cost-effective SRF accelerator, it will require the development of ultra-high-quality Nb$_3$Sn as a replacement for Nb \cite{SunRef35,SunRef31}. The reasons include Nb$_3$Sn's high critical temperature ($T_\mathrm{c}$), reduced surface resistance, enhanced quality factors, avoidance of complex cooling requirements, and the potential for large accelerating gradients. Besides small-scale applications, these properties of Nb$_3$Sn also significantly contribute to large-scale accelerators by savings in size, cost, and energy consumption while offering new capabilities to meet high demands from cutting-edge research. Additionally, improving Nb$_3$Sn material properties may enable its use in the emerging SRF cavity-based quantum computing that has shown ultra-high coherence time ($\sim$\,2\,s for Nb\cite{SunRef17,SunRef28,SunRef26}).

\begin{figure}[tbp]
\centering
\includegraphics[width=\linewidth]{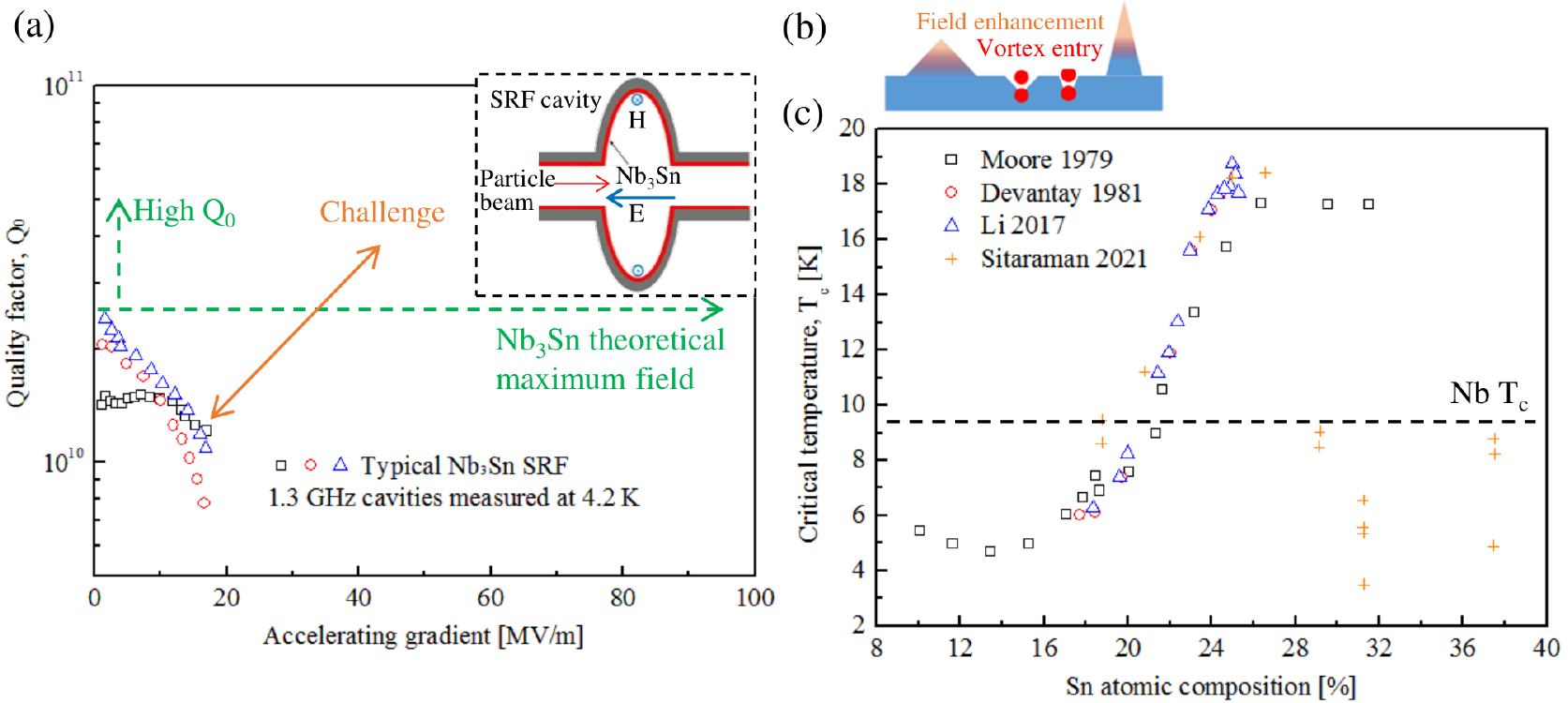}
\caption{(a) Comparison of RF performance between experimental results (vapor-diffused cavities measured at 4.2\,K) \cite{SunRef18} and theoretical predictions \cite{SunRef19} in Nb$_3$Sn SRF resonant cavities. The insert shows the schematic of a cavity. (b) Schematics showing multiple issues induced by surface roughness, \textit {e.g.}, field enhancement and vortex nucleation. (c) $T_\mathrm{c}$ variation induced by non-stoichiometry in Nb$_3$Sn (adapted from measured \cite{SunRef15,SunRef16} and calculated \cite{SunRef9,SunRef10} data).}
\label{FigS27}
\end{figure}

Nb$_3$Sn is expected to support large accelerating gradients ($E_\mathrm{acc}$) of up to 100\,MV/m, owing to the high predicted superheating field ($\sim$\,400\,mT \cite{SunRef19}) that doubles Nb's value ($\sim$\,200\,mT). The $T_\mathrm{c}$ of Nb$_3$Sn (18\,K) is notably higher than that of Nb (9\,K), which reduces surface resistance ($R_\mathrm{s}$) and thus boosts quality factors ($Q_\mathrm{0}$) following $Q_\mathrm{0}\,\propto\,1\,/\,R_\mathrm{s}$. $R_\mathrm{s}$ is the sum of residual resistance ($R_\mathrm{0}$) and temperature-dependent BCS resistance ($R_\mathrm{BCS}$) that exponentially depends on $T_\mathrm{c}$ ($R_\mathrm{BCS}\,\propto\,1\,/\,T\,\times\,\mathrm{exp}\,(-\,T_\mathrm{c}\,/\,T)$). Table~\ref{SunTable} shows the reduction of BCS resistances by replacing Nb with Nb$_3$Sn, especially at high operation temperatures (4.2\,K). Also, the higher $T_\mathrm{c}$ allows for the replacement of cooling sources from costly, specialized, complex helium cryogenics (2\,K operation) and cryomodules to commercial cryocoolers (4\,K operation) \cite{SunRef34,SunRef106,SunRef107,SunRef108}.

\begin{table}[bp]
\centering
\caption{Comparison of measured BCS resistances at multiple operation temperatures at the 1.3\,GHz frequency and low RF fields for typical Nb, nitrogen-processed Nb, vapor-diffused Nb$_3$Sn, and electrochemically synthesized Nb$_3$Sn. The values for 2\,K Nb$_3$Sn only reflect consistent comparisons, with uncertainties.}
\begin{tabular}{ |p {3cm} | p {3cm}| p {3cm} | p {3cm} | p {3cm} |} 
  \hline
  Operation temperature & Typical Nb & Typical nitrogen-processed Nb & Typical vapor-diffused Nb$_3$Sn & Electrochemically synthesized Nb$_3$Sn (this work)  \\ 
   \hline
  10\,K & - & - & 500\,--\,700\,n$\Omega$\cite{SunRef74} & 300\,--\,400\,n$\Omega$ \\ 
  \hline
  4.2\,K & 900\,n$\Omega$ & 450\,n$\Omega$ & \,6\,--\,10\,n$\Omega$ \cite{SunRef18,SunRef74} & $\sim$\,3\,n$\Omega$\\ 
  \hline
  2\,K & 15\,n$\Omega$ & 8\,n$\Omega$ & $\sim$\,1\,n$\Omega$ \cite{SunRef18,SunRef74} & $<$\,1\,n$\Omega$\\ 
  \hline
\end{tabular}
\label{SunTable}
\end{table}

\begin{figure}[b!]
\centering
\includegraphics[width=17 cm]{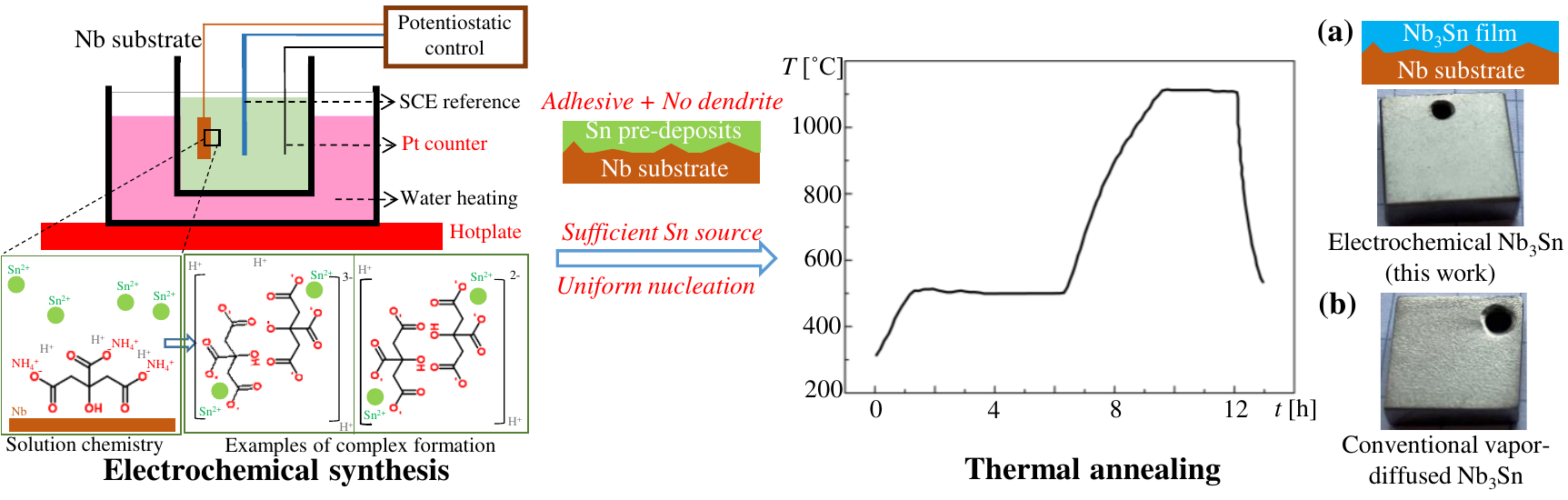}
\caption{\textbf{Process flow combines electrochemistry and post-annealing to achieve smoothness, stoichiometry, and purity in Nb$_3$Sn.} The inserts show pictures of (a) electrochemically synthesized and (b) vapor-diffused Nb$_3$Sn.}
\label{SunFig1}
\end{figure}

However, surface roughness \cite{SunRef3,SunRef4,SunRef6,SunRef18,SunRef35,SunRef38,SunRef39,SunRef79}, non-stoichiometry \cite{SunRef3,SunRef4,SunRef8,SunRef9,SunRef10,SunRef15,SunRef16,SunRef18,SunRef35,SunRef38,SunRef39,SunRef40}, and impurities \cite{SunRef3,SunRef4,SunRef10,SunRef18,SunRef35,SunRef40} are the three top-priority issues in Nb$_3$Sn. Surface roughness excessively enhances local magnetic fields \cite{SunRef6} and causes premature quenches, \textit {i.e.}, the loss of superconductivity at elevated fields \cite{SunRef39} (Fig.~\ref{FigS27}a,b). Also, roughness and local variations enable early magnetic vortex nucleation (Fig.~\ref{FigS27}b) \cite{SunRef3}. Prior to this work, vapor diffusion was the only successful process to make Nb$_3$Sn SRF cavities \cite{SunRef35,SunRef18}. Vapor-diffused Nb$_3$Sn films \textit{typically} exhibit an average surface roughness ($R_\mathrm{a}$) of above 300\,nm (\textit{e.g.}, over 20\,$\times$\,20\,$\mu$m$^2$ areas) and maximum peak-to-valley heights above 2\,$\mu$m, highlighting the importance of achieving smoother film surfaces \cite{SunRef38,SunRef40,SunRef18,SunRef35,SunRef39,SunRef74,SunRef79}. The adoption of thinner films suggests a roughness reduction; nevertheless, the reproducibility of their RF performance has been an issue, probably due to their excessive thinness (versus the field penetration depth) and subsequent interface issues \cite{SunRef31}. Furthermore, stochiometric variations dramatically degrade $T_\mathrm{c}$ and induce local heating; for example, a 3\,at.\% variation reduces $T_\mathrm{c}$ by 50\% (Fig.~\ref{FigS27}c) \cite{SunRef8,SunRef9,SunRef10,SunRef15,SunRef16}. Also, impurities strongly affect the electron mean free path and hence surface resistance. In vapor-diffused Nb$_3$Sn, Sn-deficient grains and impurities have been observed at the local surface regions with larger dimensions than the $\sim$\,3\,nm coherence length \cite{SunRef38,SunRef40,SunRef105}. These crystal defects limit the minimization of BCS resistance and the enhancement of quality factors (Table~\ref{SunTable}). It is thus essential to investigate methods for reducing roughness, attaining stoichiometry, and controlling impurities in Nb$_3$Sn, along with surface engineering at the scale of hundreds of nanometers dictated by the field penetration depth.  

Studies on vapor-diffused Nb$_3$Sn \cite{SunRef9,SunRef38,SunRef40} suggest that these material issues arise from insufficient and non-uniform Sn-vapor supply. Local variations in Sn flux during vapor diffusion result in an uneven spatial distribution of growth rates, attributing to roughness. Likewise, inadequate Sn sources and slow supply rates produce initially Sn-deficient grains ("bad" Nb$_3$Sn) following the 18\,at.\%-Sn boundary in the phase diagram; subsequently, slow Sn diffusion within these grains kinetically limits their conversion into stoichiometric "good" Nb$_3$Sn \cite{SunRef73,SunRef48,SunRef38,SunRef9}. Thus, we intend to design a deposition process that ensures a sufficient amount and rate of Sn supply during nucleation. 

As illustrated in Fig.~\ref{SunFig1}, we deposit a prerequisite Sn source on the Nb surface using our electrochemical recipe and subsequently convert the film to Nb$_3$Sn through thermal annealing. This process yields smooth and stoichiometric Nb$_3$Sn films ($R_\mathrm{a}$\,=\,54\,$\pm$\,4\,nm over 20\,$\times$\,20\,$\mu$m$^2$ areas) on the industry-standard Nb substrate (initial roughness $>$\,100\,nm), along with minimal impurity concentrations of hydrogen (H), carbon (C), oxygen (O), and nitrogen (N). Here, we present a detailed investigation of Sn-surfactant electrochemistries on Nb and the design principles of how initial electrochemical pre-deposits affect the material properties of resulting Nb$_3$Sn films. We provide the atomic, chemical, structural, surface, and superconducting properties of electrochemically synthesized Nb$_3$Sn, and compare them with vapor-diffused films. We demonstrate excellent RF performance, \textit {e.g.}, ultra-low BCS resistances and high quality-factors, of the first electrochemically synthesized Nb$_3$Sn cavity as a proof of concept for use in SRF cavities and other superconducting devices.

\section{Method}

\subsection{Electrochemical synthesis and characterization}

The industry-standard Nb substrates with superconducting Residual Resistivity Ratios (RRR) greater than 300 were mechanically polished and electropolished using a mixture of 9 parts 98\% sulfuric acid (H$_2$SO$_4$) and 1 part 48\% hydrofluoric acid (HF), resulting in surface roughness values of $R_\mathrm{a}$\,=\,$\sim$\,100\,nm and absolute-maximum $R_\mathrm{z}$\,=\,$\sim$\,2.3\,$\mu$m (Table~S4). Nominal substrate oxides may persist after HF cleaning before Sn pre-deposition.

Sn electrochemical deposition was performed using a Princeton Applied Research VersaSTAT 3-500 potentiostat system. As illustrated in Fig.~\ref{SunFig1}, a three-electrode setup was employed with the Pt wire counter and saturated calomel (SCE) electrodes. A deionized water bath achieved uniform heating of reaction mixtures for sample-scale studies, and digital feedback control monitored the temperature.

SnCl$_2$ and ammonium citrate tribasic surfactant were purchased from Sigma-Aldrich and used as received. The pH of the as-prepared 0.2\,M SnCl$_2$ / 0.3\,M citrate was 4.14, while changing the SnCl$_2$ / surfactant ratio altered pH values as monitored by pH meters (Fig.~S2). pH values were also adjusted by adding HCl or NaOH in the optimum baths, covering the range of 0 to 14. Solutions were stirred for 5\,min after dissolving the chemicals and let stand for another 5\,min. (White clouds were observed on mixing, likely due to low-solubility SnCl$_2$ that later dissolved on standing to obtain clear reaction solutions or Sn(OH)Cl produced from the reaction between SnCl$_2$ and H$_2$O in low-pH baths.) 

Cyclic voltammetry (CV) measurements were performed to determine Sn electrochemistries on Nb at varied scan rates (20\,--\,100\,mV/s). Effects of temperature (30\,--\,90\,°C), pH (0\,--\,14), surfactant (0.2\,--\,3.7\,M) and precursor (0.2\,--\,0.7\,M) concentrations, redox potentials, substrate oxide thicknesses, and stirring conditions along with surfactant- and precursor-only control studies were systematically investigated. 

A potentiostat controlled the deposition. The deposition time was altered from 2.5 to 20\,min. After deposition, samples were cleaned with methanol, dried, and sealed in plastic bags. The deposition processes were repeated in 22 batches with multiple samples in each set.

\subsection{Post thermal annealing}

The electrochemical pre-deposits were annealed in a high-vacuum (10$^{-6}$\,Torr) furnace for conversion to Nb$_3$Sn. As shown in Fig.~\ref{SunFig1}, samples were heated at 500\,\textdegree C for 5\,h for nucleation and subsequently at 1100\,\textdegree C for 3\,h for alloying, followed by slow cooling in the furnace without additional control. For comparison, vapor-diffused samples were prepared using the same heating profile \cite{SunRef18}. 

\subsection{Surface, chemical, atomic, and structural characterizations}

The electrochemical pre-deposits and converted Nb$_3$Sn films were imaged by a Zeiss Gemini scanning electron microscope (SEM) equipped with an in-lens detector under the low voltages (1\,--\,5\,kV) to evaluate the film uniformity, surface morphology, grain information, and dendrite formation. Films cross-sections were polished and scanned using Thermo Fisher Helios G4 UX focused ion beam (FIB) to determine the film thicknesses and prepare the scanning transmission electron microscopy (STEM) specimens. Surface profiles were examined and quantified using Asylum MFP-3D atomic force microscopy (AFM), with an analysis area of 20\,$\times$\,20\,$\mu$m$^2$. Cross-sectional microscopy images provided supporting evidence of surface smoothness. 

A combination of techniques, including energy-dispersive X-ray spectroscopy (EDS) under SEM and STEM microscopies, X-ray photoelectron spectroscopy (XPS), and secondary ion mass spectrometry (SIMS), were utilized to determine the chemical composition, stoichiometry, and impurity information. Cross-sectional EDS/STEM provided composition mappings, and stoichiometry depth profilings were achieved by XPS combined with ion etching, with the etching rate calibrated using actual film thicknesses. For the XPS survey scans, we used the SSX-100 XPS instrument, and the stoichiometry values obtained from high-intensity survey scans are reliable. Monochromatic Al K$_{\alpha}$ X-ray (1486.6\,eV) photoelectrons were collected under a 10$^{-9}$\,Torr vacuum from an 800\,$\mu$m analysis spot with a 55\textdegree~emission angle. The scan parameters were set to 150\,eV pass energy, 1\,eV step size, and 100\,s/step. For depth profiles, a 4\,kV Ar$^+$ beam with a spot size of $\sim$\,5\,mm was rastered over a 2\,$\times$\,4 mm$^2$ area. For the high-resolution XPS spectra, we used the PHI Versaprobe XPS instrument with a 100 $\mu$m monochromatic Al K$_{\alpha}$ X-ray (1486.6\,eV) beam. The scan parameters were set to a 45\textdegree~emission angle, 26\,eV pass energy, 50\,ms/step, and up to 60 sweeps with the dual-neutralization on. For depth profiles, a 3\,keV Ar$^+$ beam was rastered over 2\,$\times$\,2\,mm$^2$ area with Zalar rotation. The SIMS data was provided by Eurofins EAG Laboratories.

Cross-sectional phase and grain mappings that resolve high contrasts were obtained by FEI F20 4D-STEM equipped with a high dynamic range EMPAD (electron microscope pixel array detector) and analyzed by the EWPC (exit-wave power cepstrum) transform \cite{SunRef47}. High-resolution Rigaku SmartLab X-ray diffraction (XRD) was measured to support the determination of phase and grain orientations. The X-ray generated by a Cu target was converted into a parallel, monochromatic beam and directed onto the sample surface after passing through a 5\,mm divergence slit. The K$_{\alpha}$ signal with a wavelength of 0.154\,nm was collected by filtering the diffracted beam. The parallel slit analyzer and 5\textdegree~Soller slits were used to resolve high-resolution signals. The 2$\theta$ scan was performed with a step size of 0.05\textdegree. Atomic imaging was obtained using FEI/Thermo Fisher Titan Themis STEM. Additionally, tape tests were performed to confirm the adhesion strength for electrochemical pre-deposits on Nb.

\subsection{Superconducting property and RF performance}

$T_\mathrm{c}$ values were determined by (1) temperature-dependent resistivity measurements under different DC fields using a Quantum Design Physical Property Measurement System (PPMS) under the AC transport mode and (2) flux expulsion tests while warming up the cryostat from liquid-helium temperature. For the resistivity measurements, the sample surface was wedge-bonded to the sample puck using four 25\,$\mu$m aluminum wires (West Bond 747630E); the fields (0\,T and 0.5\,T) were applied perpendicular to the sample surface. During the flux expulsion tests, we utilized two flux gate magnetometers positioned at different locations on the electrochemically synthesized cavity, oriented parallel to its axis. As the superconductor undergoes warming up, it transitions from the superconducting state to the normal conducting state, causing previously expelled magnetic flux to re-enter the superconductor. This results in an abrupt change in the magnetic fields in the vicinity of the cavity. The cryostat is magnetically shielded such that the dominant component of the magnetic fields in the cryostat aligns along the cavity's central axis with a magnitude $<$\,10\,mG. We recorded either an increase or a decrease in the magnetic fields depending on the orientation to indicate the $T_\mathrm{c}$ (For more information about this measurement setup, see Ref. [\cite{SunRef110,SunRef111}]). 

To evaluate the RF performance, we scaled up the electrochemical and annealing process to the large sizes at the inner surface of a 1.3\,GHz SRF cavity. The cavity was electropolished with a nominal 100\,$\mu$m removal. The deposition was performed in an inert gas glovebox with O$_2$ and H$_2$O levels below 0.5\,ppm. To enhance large-scale deposition and streamline glovebox operation, we improved the heating system by replacing the water bath with wrapped fiberglass heating tape. We positioned two thermocouples at different locations on the cavity to monitor the plating temperature. The electrochemical and annealing parameters were set the same as with sample-scale depositions; nevertheless, the plating current density was low, likely due to thin electrodes used for deposition on a large surface area ($\sim$\,0.2\,m$^2$). $R_\mathrm{s}$'s were measured at 1.6\,--\,10\,K temperatures under increasing RF fields. Temperature-dependent $R_\mathrm{BCS}$'s were determined by the difference between total $R_\mathrm{s}$ and an estimate of $R_\mathrm{0}$ (the average of $R_\mathrm{s}$'s at 2.4\,--\,2.6\,K temperatures).


\section{Sn electrochemistry on Nb: effects of surfactant, temperature, pH, concentration, and substrate surface oxide}

\begin{figure}[b!]
\centering
\includegraphics[width=\linewidth]{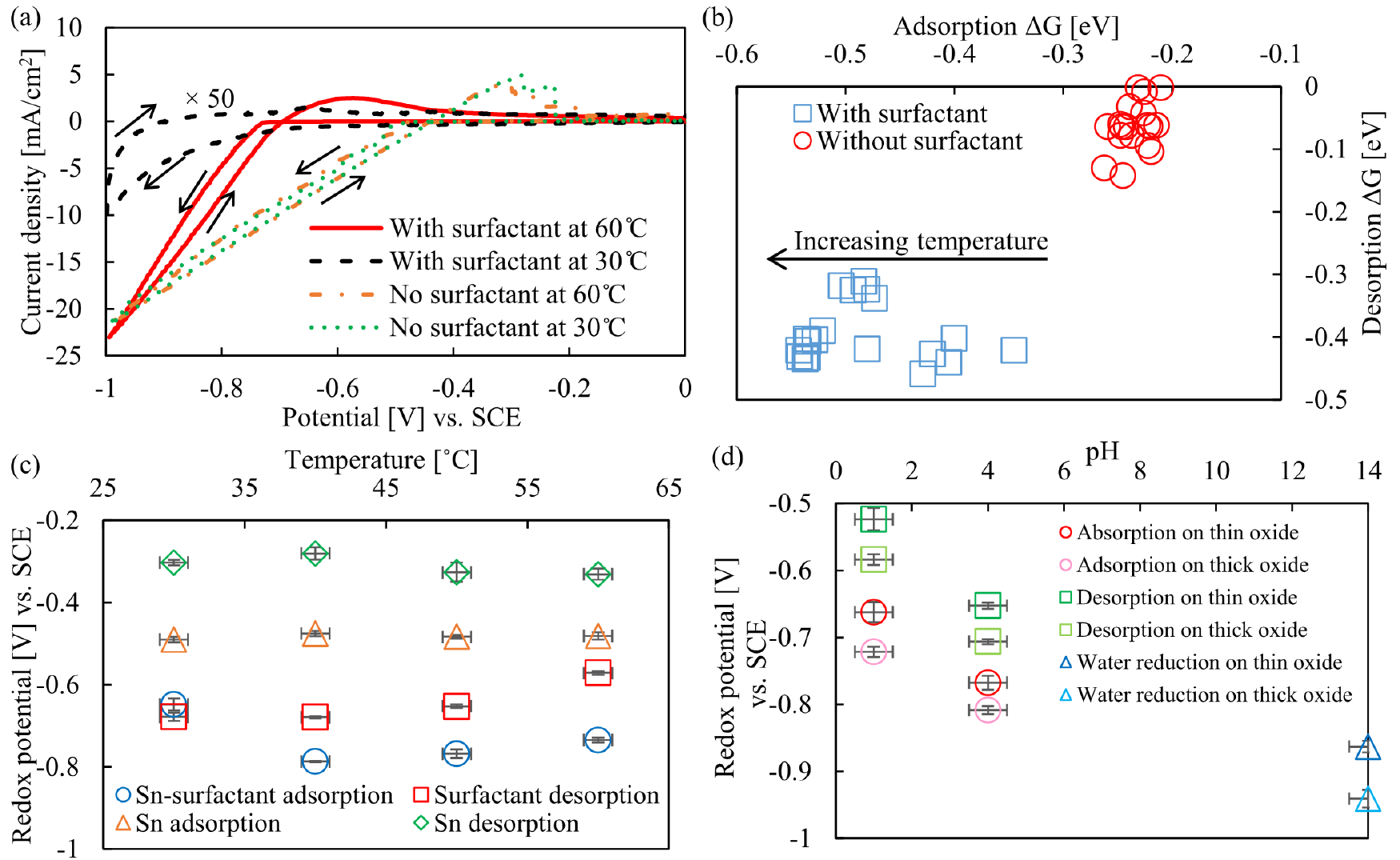}
\caption{\textbf{Electrochemical analysis of Sn plating on the Nb surface.} (a) CVs of Sn and Sn-surfactant electrochemistries on Nb scanned at a rate of 40\,mV/s. (b) Effect of surfactant on adsorption and desorption half-cell free energies ($\Delta\,G$). (c,d) Effects of bath (c) temperature and (d) pH on redox potentials for surface adsorption/desorption on (c,d) thin and (d) thick native Nb oxides.}
\label{FigS26}
\end{figure}

A central prerequisite for achieving stoichiometry and low roughness in converted Nb$_3$Sn is to ensure the high quality of Sn pre-deposits, $\textit{i.e.}$, uniform, smooth, dendrite-free, and adhesive. We have developed an electrochemical recipe for Sn pre-deposition, featuring a manufacturing solution (low-cost, scalable, selective, and reactive) for versatile depositions on intricate, curved structures, including large-sized (several centimeters to meters) accelerating cavities and small-sized (several nanometers to micrometers) quantum devices. 

To date, direct Sn electrodeposition on Nb is missing. The challenges include unwanted dendrite formation (abnormal Sn vertical growth due to low lateral surface diffusion) and peel-off issues (poor adhesion at the Sn/Nb interface). Dendrites disrupt film uniformity and increase the resulting roughness of Nb$_3$Sn. The low adhesion impacts any required handling or post-treatment processes, $\textit{e.g.}$, high-pressure (up to 140\,bar) water rinsing used for SRF cavity preparation. Poor adhesion could also pose issues with thermal transport, \textit{i.e.}, RF dissipation to cooling on the cavity outside. 

Cu, Au, Pt, and Ti seed layers \cite{SunRef50,SunRef51,SunRef52}, as well as bronze (Sn-Cu alloy) / Nb extraction \cite{SunRef54}, can improve adhesion and uniformity. However, these indirect methods inevitably introduce normal-conducting contaminations (\textit{e.g.}, Cu inclusions and bronze byproducts), causing severe thermal runaway and immediate quench for RF use. Attempts at electrodeposition approaches using Cu \cite{SunRef50,SunRef77} or bronze \cite{SunRef54,SunRef78} seed layers have been explored for SRF cavities. In the Sn/Cu/Nb approach \cite{SunRef50,SunRef77}, Cu seeds are first electroplated on Nb, followed by Sn films electroplated on Cu, and the film stack is annealed to generate a bronze/Nb$_3$Sn structure. Etching is required to remove the surface bronze, but controlling this etching process is undesirable. Another issue is the presence of performance-degrading Cu inclusions appearing inside Nb$_3$Sn, which hinders further development of this process for SRF use\cite{SunRef77}. Similarly, in the bronze/Nb approach, a bronze layer is directly electroplated on Nb, followed by annealing, resulting in similar challenges, \textit{i.e.}, bronze residues and Cu inclusions \cite{SunRef54,SunRef78}.   

Here, we consider an alternative electrochemistry that leverages surfactants to enable a seed-free process. The addition of surfactants, such as citrate \cite{SunRef51,SunRef52,SunRef56}, gluconate \cite{SunRef57,SunRef58,SunRef59}, tartrate \cite{SunRef61}, pyrophosphate\cite{SunRef62}, and other surface-active additives\cite{SunRef63,SunRef64}, has been investigated on Cu, Au, and steel surfaces. However, the Sn-unfriendly Nb substrate and substrate oxides, associated with solution chemistries, temperature, and pH, significantly alter the Sn growth mechanism and deposit quality. For example, reported surface morphologies \cite{SunRef57,SunRef59,SunRef62,SunRef63} showed incomplete films with large surface roughness. Due to the nanoscale sensitivity to multiple surface effects for SRF use, we should best optimize the Sn electroplating conditions on the Nb surface.

The best deposits were determined from a 0.2\,M SnCl$_2$ aqueous bath containing 0.3\,M ammonium citrate tribasic surfactant at temperatures above 50\,\textdegree C, plated at a potential of -\,0.7 to -\,0.8\,V versus the saturated calomel electrode (SCE), without additional acid/base and stirring.
Fig.~\ref{FigS26}a illustrates cyclic voltammetries (CV) of Sn-surfactant versus Sn$^{2+}$ on Nb at different bath temperatures. The presence of surfactants alters the chemistry of the plating solution. In the absence of surfactants, regardless of the plating temperature, the CV curves behave similarly, with white Sn(OH)Cl clouds forming before plating due to reactions between SnCl$_2$ and water. At a potential of -\,0.5\,V, the Sn$^{2+}$ reduction onset occurs as a crossover during the negative-to-positive scan, inferring Sn nucleation \cite{SunRef65,SunRef66}. Representative crossover behavior is depicted in Fig.~S1, indicating nucleation-induced overpotential, where rapid nucleation limits mass transport and discourages ion diffusion during inverted potential scanning. An oxidation peak assigned to Sn desorption emerges at -\,0.3\,V, with hydrogen generation occurring at -\,0.24\,V. The deposits are observed as SnCl$_x$ particles under SEM/EDS (Fig.~S4). 

In contrast, the addition of surfactants stabilizes the dissolution of Sn$^{2+}$ in deionized water, ending with a clear solution that generates Sn-surfactant complexes (examples shown in Fig.~\ref{SunFig1}). The nearly zero current at 30\,\textdegree C indicates a deceleration of adsorption of complexes toward the cathode, while nucleation crossovers vanish. Inactive complexes likely provoke this inhibiting behavior of reduction, \textit{e.g.}, [Sn$_2$($C_6$$H_5$$O_7$)$_2$H$_{-2}$]$^{4-}$. Upon heating to 60 \textdegree C, in contrast to surfactant-free baths, a new reducing reaction begins around the potential of -\,0.8\,V, consistent with metal complex reduction. An oxidation peak exists at -\,0.6\,V potential, where mass transport of the desorbed surfactant maximizes. 

Existing complexation chemistry\cite{SunRef67} suggests that the main species in our solution, as indicated by pH, is [Sn$_2$($C_6$$H_5$$O_7$)$_2$H$_{-1}$]$^{3-}$, along with a small amount of [Sn$_2$($C_6$$H_5$$O_7$)$_2$]$^{2-}$ and [Sn$_2$($C_6$$H_5$$O_7$)$_2$H$_{-2}$]$^{4-}$. The main pathways for our electrochemical synthesis involve the adsorption of Sn-citrate complexes and subsequent desorption of citrate ions ([($C_6$$H_5$$O_7$)H$_{2}$]$^{-}$), generating acetoacetate (($C_4$$H_5$$O_3$)H)\cite{SunRef68}, which later decomposes to acetone ((C$H_3$)$_2$CO) under heat.

Adsorption: \ce{[Sn$_2$($C_6$$H_5$$O_7$)$_2$H$_{-1}$]$^{3-}$ + 5H$^{+}$ + 4e$^{-}$ $\rightarrow$ 2Sn + 2[($C_6$$H_5$$O_7$)H$_{2}$]$^{-}$};

Desorption: \ce{[($C_6$$H_5$$O_7$)H$_{2}$]$^{-}$ $\rightarrow$ ($C_4$$H_5$$O_3$)H + H$^{+}$ + 2CO$_2$ + 2e$^{-}$}, \ce{($C_4$$H_5$$O_3$)H ->[Heat] (C$H_3$)$_2$CO + CO$_2$}.  

Using Nernst's equation (free energy $\Delta G\,\propto\, E_\mathrm{0}$ versus hydrogen electrode potential), we calculated half-cell free energies for adsorption and desorption to understand the effect of surfactant (Fig.~\ref{FigS26}b). The system favors adsorption in the presence of surfactant, which is activated by increasing temperature, whereas similar free energies for surfactant-free baths do not permit effective Sn deposition.

Fig.~\ref{FigS26}c compares the redox potentials of different processes at increasing bath temperatures. The Sn$^{2+}$ adsorption and desorption (Sn$^{2+}$\,+\,2e$^{-}$\,<-->\,Sn) are both active at low redox potentials and independent of bath temperature. Increasing temperature facilitates Sn-surfactant adsorption and surfactant desorption. The complex reactions involve multiple intermediate steps catalyzed by heat, $\textit{e.g.}$, acetoacetate decomposition. The desorption shifting is steered by the removal of products following the relation, $E_\mathrm{des}$\,$\propto$\,$T$*ln($a_\mathrm{pro}$\,/\,$a_\mathrm{rea}$), where $a_\mathrm{pro}$ and $a_\mathrm{rea}$ are activities of oxidizing products and reagents, respectively. Thus, we have chosen higher bath temperatures ($\textit {e.g.}$, 80\,\textdegree C) to compensate for any heat loss in the large-scale deposition for SRF cavities. 


Independent pH studies (pH\,=\,0\,--\,14) were conducted by adding HCl or NaOH to the optimum bath. In acidic baths, pH shifts the redox potentials as protons are involved in both adsorption and desorption reactions (Fig.~\ref{FigS26}d). The plating results varied (Table~S2), likely due to aggressive vertical growth occurring with the significantly high reduction current at pH\,=\,1 (Fig.~S11). The data suggests that a balance of rates between multiple processes (adsorption, desorption, lateral growth, vertical growth) is needed to ensure a layer growth mode. (The basic baths react with SnCl$_2$ and produce a cloudy white solution; CVs only show water reduction.) 

Native oxides on the Nb substrate affect the plating results. Successful depositions were achieved by removing the native oxides using HF before initiating any plating process (although thin oxides may nominally exist when using regular chemical hoods). CVs (Fig.~\ref{FigS26}d) revealed that thick native oxides ($\sim$\,7\,nm) shift the redox potentials, and film formation was not observed on the thick oxides regardless of the pH conditions. After controlling the oxides to zero nominal thinness through HF pre-soaking, we obtained high-quality deposits in both regular chemical hoods and the inert-gas glovebox. However, the glovebox is preferred for achieving stable results. Therefore, we performed the 1.3\,GHz SRF cavity deposition in the glovebox.

The Supplementary Information details the optimization of synthesis and characterization of pre-deposits.

\begin{figure}[b!]
\centering
\includegraphics[width=\linewidth]{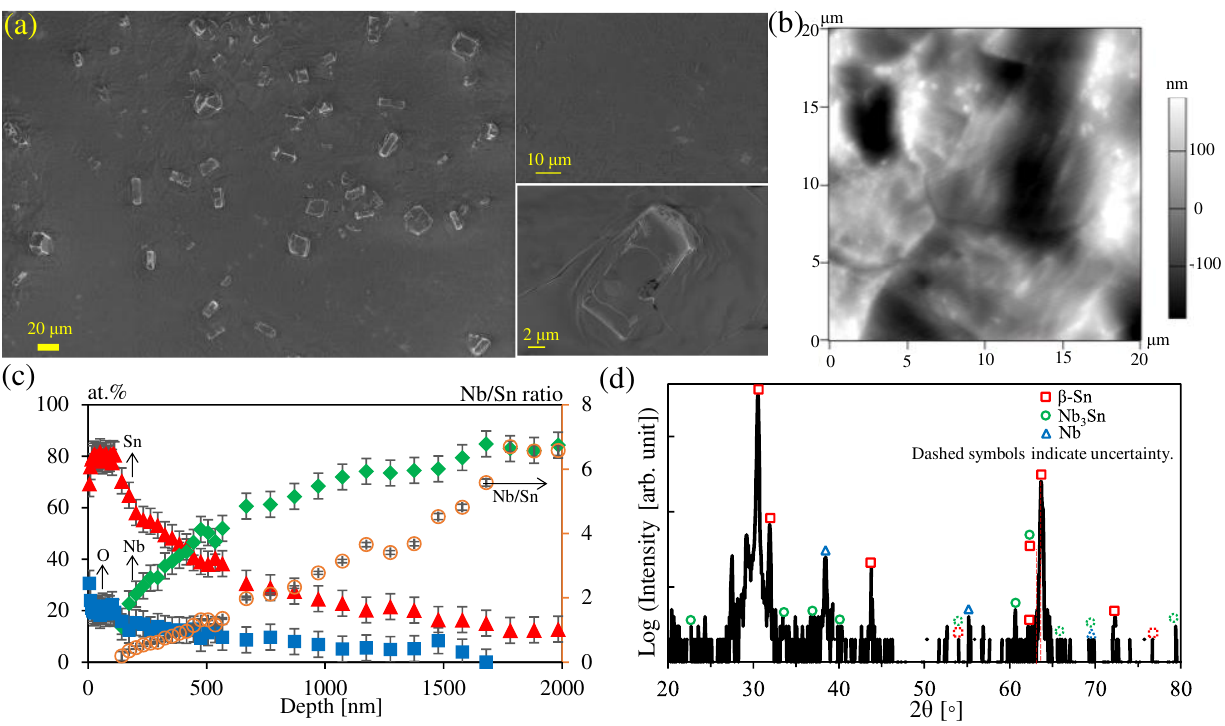}
\caption{\textbf{Characterization of Sn pre-deposits plated for 10\,min.} (a) SEM and (b) AFM images showing smooth films with embedded crystals (see inserts). (c) XPS depth profiling of Sn, Nb, and O. (d) XRD pattern showing a strong $\beta$-Sn preference and some Nb$_3$Sn diffractions.}
\label{FigS28}
\end{figure}

\section{Characterization of optimized electrochemical pre-deposits}

Fig.~\ref{FigS28} shows the surface, chemical, and structural properties of 10 min pre-deposits made from the optimized electrochemical condition. SEM confirms a uniform and smooth film with no dendrites formed. AFM mapping in Fig.~\ref{FigS28}b determines $R_\mathrm{a}$ as low as 50\,nm. Some embedded "island" crystals with a size of $\sim$\,5\,$\mu$m exist. (This behavior is not further investigated since the converted Nb$_3$Sn did not exhibit any particular feature on these "island" sites.) After FIB polishing on the film edge, the cross-sectional SEM reveals a 3.3\,$\mu$m thickness over 10\,min deposition, corresponding to a growth rate of 0.3\,$\mu$m/min. The optimized electroplating conditions have been reproducibly demonstrated. Moreover, tape tests showed excellent adhesion, and the peel-off issue has been resolved, with plated films on the SRF cavity surviving the high-pressure water rinse. 



By quantifying the atomic concentrations of Sn, Nb, and O using XPS (Fig.~\ref{FigS28}c) and analyzing their peak positions and FWHM (full width at half maximum) values (Fig.~S14) as a function of depth, we find that the film consists of two distinctive regions. The top region (0\,--\,150\,nm) comprises metallic Sn and SnO$_x$. The large FWHM values of O and Sn indicate the convolution of multiple motifs, with the Sn peak shifting from 487.5\,eV to $\sim$\,485\,eV and oxygen shifting from 530.6\,eV to 530\,eV. The base region ($\sim$\,150\,nm\,--\,1.6\,$\mu$m) shows a gradient of Nb/Sn ratios. FWHM values remain nearly constant for all elements, with Sn and Nb peaks shifting by $\sim$\,0.5\,eV and O persisting at 530\,eV. These low binding energies align with the positions of metallic Nb and Sn.


XRD diffractions (Fig.~\ref{FigS28}d) exhibit prominent peaks consistent with body-centered tetragonal $\beta$-Sn and cubic substrate Nb. There may exist some low-crystallinity Nb$_3$Sn. Despite the high-resolution capability of Rigaku XRD (able to resolve <0.01$^{\circ}$ resolution), the identification of certain peaks is challenging due to their low intensity. Using an empirical method, we collected data at a small step size of 0.05$^{\circ}$ and a low scan rate of 1.44\,$^{\circ}$/min, treating the global noise across the entire 2$\theta$ scan as the background. This method, with caution, identifies nearly all diffractions from A15 Nb$_3$Sn, along with additional diffractions from $\beta$-Sn and bcc Nb (as referenced in Table~S3). Notably, strong diffractions at 27.5$^{\circ}$ and 29.2$^{\circ}$ remain unindexed with any possible phases for Sn, Nb, Nb$_3$Sn, and SnO$_2$. Further investigations are needed to identify local phases after the electrochemical synthesis. This material system is fully converted to A15 Nb$_3$Sn after thermal annealing.

\section{Correlation between Sn pre-deposition and Nb$_3$Sn surface roughness}

\begin{figure}[b!]
\centering
\includegraphics[width=\linewidth]{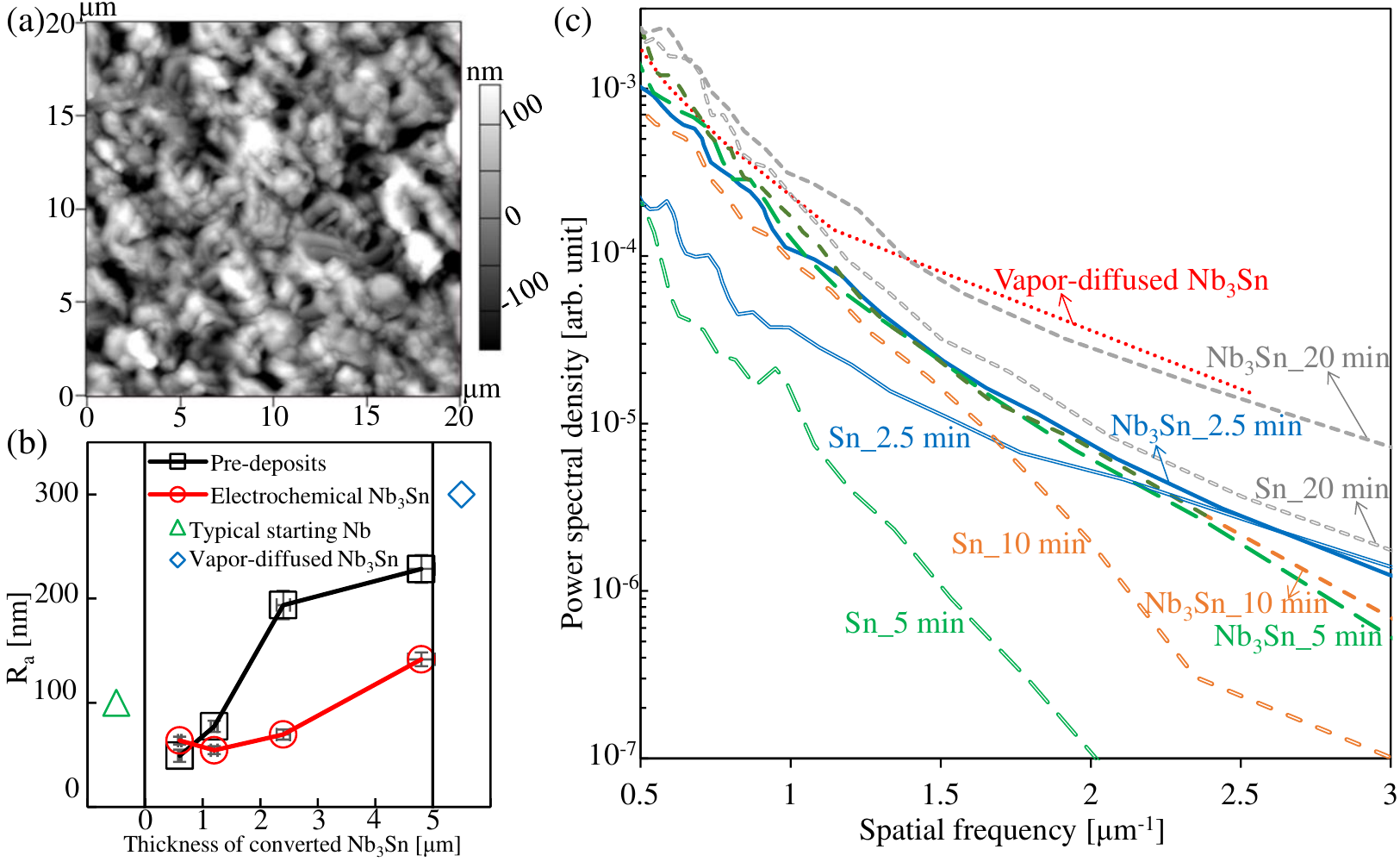}
\caption{\textbf{Effect of Sn pre-deposition on resulting Nb$_3$Sn roughness.} (a) AFM surface morphology of electrochemically synthesized Nb$_3$Sn converted from 2.5\,min pre-deposition. (b) $R_\mathrm{a}$ roughnesses of converted Nb$_3$Sn versus their film thicknesses, compared with the corresponding pre-deposits, typical $\sim$\,3\,$\mu$m thick vapor-diffused Nb$_3$Sn \cite{SunRef18,SunRef39,SunRef74,SunRef79}, and Nb substrates. (c) PSD obtained through FFT algorithms for electrochemically synthesized Nb$_3$Sn and pre-deposits with varying plating times (2.5\,--\,20\,min), compared with vapor-diffused Nb$_3$Sn. The labels, for example, Sn\_2.5\,min and Nb$_3$Sn\_2.5\,min, refer to the Sn pre-deposit and converted Nb$_3$Sn samples obtained after a 2.5\,min electrochemical deposition of Sn, respectively.}
\label{SunFig2}
\end{figure}


By employing chronoamperometry at a 0.3\,$\mu$m/min plating rate, we controlled the pre-deposit thickness, ranging from 800\,nm to 6.6\,$\mu$m, as determined by cross-sectional SEM following FIB polishing. Subsequently, the films were subjected to vacuum annealing using the heating profile depicted in Fig.~\ref{SunFig1} to convert them into Nb$_3$Sn. The AFM scan in Fig.~\ref{SunFig2}a illustrates the Nb$_3$Sn surface topography, exhibiting $\sim$\,1\,$\mu$m grain-like structures converted from the 2.5\,min pre-deposition (AFM data for all conditions in Fig.~S15). Unlike vapor-diffused films characterized by surface variations, ranging from thin to abnormally large grains \cite{SunRef38}, the electrochemically synthesized Nb$_3$Sn shows uniformly distributed small grains (SEM images in Fig.~S17).
  
A visual comparison in Fig.~\ref{SunFig1} between electrochemically synthesized and vapor-diffused Nb$_3$Sn undoubtedly demonstrates the smooth surface converted from electrochemical pre-deposits, with empirical statistical significance. In Fig.~\ref{SunFig2}b, we compare AFM quantifications of electrochemically synthesized and vapor-diffused Nb$_3$Sn using 20\,$\times$\,20\,$\mu$m$^2$ areas with Nb substrates and pre-deposits as references. The $R_\mathrm{a}$ of pre-deposits increases with longer plating time. Dendrites appear at 20\,min of plating and jeopardize the surface profile of pre-deposits (Fig.~S15g). The converted Nb$_3$Sn exhibits $R_\mathrm{a}$ values that are essentially minimized to below 100\,nm, unless dendrites were formed during pre-deposition. These values are approximately equivalent to the starting Nb $R_\mathrm{a}$ ($\sim$\,100\,nm). In contrast, the typical roughness of $\sim$\,3\,$\mu$m thick vapor-diffused Nb$_3$Sn is observed to be 5\,$\times$ higher \cite{SunRef38,SunRef18,SunRef39,SunRef74,SunRef79} compared to electrochemically synthesized Nb$_3$Sn. The smallest $R_\mathrm{a}$ value for electrochemically synthesized Nb$_3$Sn is 54\,nm, achieved during 2.5\,--\,10 min of dendrite-free plating, with variations ($<$\,10\,nm) between baths. Dendrite formation during pre-deposition causes large roughnesses of converted Nb$_3$Sn ($\sim$\,150\,nm), approaching the $\sim$\,300\,nm value found in vapor-diffused Nb$_3$Sn. Indeed, electrochemical pre-deposits, if smooth and dendrite-free, can improve the surface profile of the converted Nb$_3$Sn. Root-mean-squared ($R_\mathrm{q}$) and $R_\mathrm{z}$ roughnesses find similar trends (Table~S4 and Fig.~S16). 

Local inspections using cross-sectional STEM confirm a smooth surface on electrochemically synthesized Nb$_3$Sn (Fig.~\ref{SunFig3}d), whereas vapor-diffused Nb$_3$Sn comprises both smooth and uneven regions. The smooth regions of both vapor-diffused and electrochemically synthesized Nb$_3$Sn look similar. However, the rough regions on vapor-diffused Nb$_3$Sn, \textit {e.g.}, an extreme example in Fig.~\ref{SunFig3}e, exhibit surface variations of 1\,--\,2\,$\mu$m, along with regions devoid of the Nb$_3$Sn film. The rationale for choosing an extreme example stems from the consequence that any nano-scale defects exceeding the coherence length would completely degrade the meter-size cavity's performance. This rough surface is likely improved in vapor diffusion by a "correct" Sn-vapor flux and thinner films \cite{SunRef31}, but additional performance issues emerge, \textit{e.g.}, poor interfaces and reproducibility. Our electrochemical Sn pre-deposition remains a feasible and promising approach to collectively resolve these issues.   

To effectively evaluate surface profiles, we utilized fast Fourier transform (FFT) algorithms using AFM surface height data to quantify the power spectral densities (PSD) of surface features at different spatial frequencies (Fig.~\ref{SunFig2}c) \cite{SunRef109}. Our focus was on frequencies ranging from 0.5\,--\,3\,$\mu$m$^{-1}$, where sharp features significantly enhance local fields. Nb$_3$Sn films derived from dendrite-free pre-deposits, regardless of their plating times (2.5\,--\,10\,min), exhibit PSDs approximately 5\,$\times$ lower than those of vapor-diffused Nb$_3$Sn, \textit{e.g.}, at the 2 $\mu$m$^{-1}$ frequency. The highest PSDs for electrochemically synthesized Nb$_3$Sn are observed when dendrites form during the 20\,min plating, resembling the values found in vapor-diffused Nb$_3$Sn. Furthermore, the dendrite-free pre-deposits demonstrate even lower densities than their converted Nb$_3$Sn counterparts, suggesting that optimization of thermal annealing may further improve the surface conditions.

Our findings emphasize the significance of smooth Sn pre-deposits in overcoming spatial variations in nucleation rates and reducing the kinetic demands for Sn diffusion by offering adequate Sn sources in close proximity to the nuclei during alloying. In contrast, studies on vapor diffusion \cite{SunRef9,SunRef38} revealed that an insufficient and spatially varying Sn supply resulted in erratic nucleation; the presence of long-range obstructed lateral Sn diffusion on Nb surfaces or low Sn bulk diffusion within Nb and Nb$_3$Sn grains \cite{SunRef9,SunRef38} further exacerbated these problems. Here, the use of Sn pre-deposits avoids high surface and chemical barriers for Sn diffusion and ensures a uniform distribution of nucleation sites, thereby ameliorating the surface profile.

\section{Improved stoichiometry in electrochemically synthesized Nb$_3$Sn and comparison with vapor diffusion}

\begin{figure}[t!]
\centering
\includegraphics[width=\linewidth]{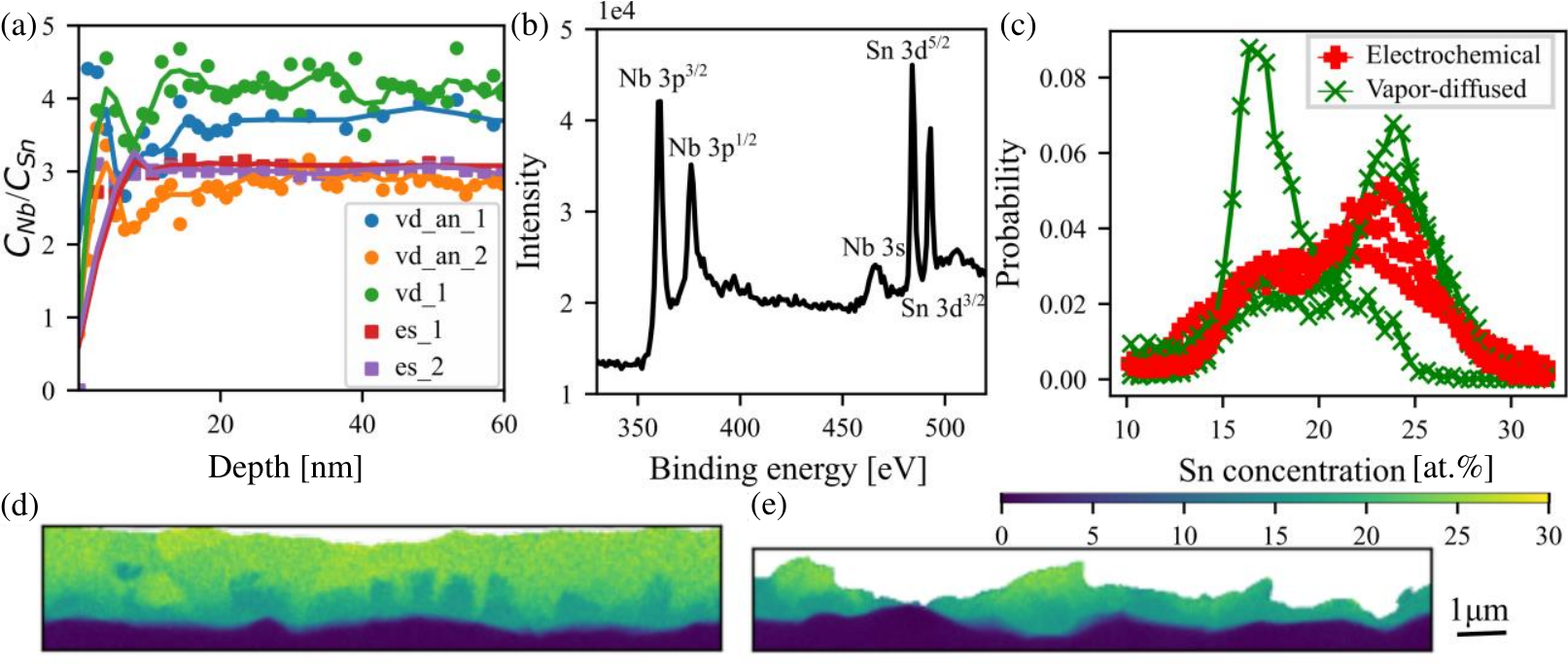}
\caption{\textbf{Comparison of stoichiometry between electrochemically synthesized and vapor-diffused Nb$_3$Sn.} (a) XPS depth profiling of Nb$_3$Sn produced by electrochemical synthesis (es) versus vapor diffusion (vd) with/without pre-anodization (an). (b) Typical XPS spectrum for electrochemically synthesized Nb$_3$Sn taken after etching surface oxides. (c) Probability of Sn concentration over the entire cutout region (\textit{e.g.}, as shown in (d,e)) analyzed by cross-sectional 4D-STEM/EDS. (d,e) Cross-sectional composition maps of (d) electrochemically synthesized Nb$_3$Sn and (e) vapor-diffused Nb$_3$Sn (an extreme example).}
\label{SunFig3}
\end{figure}

Stoichiometry within the first few hundreds of nanometers of the surface, where RF fields penetrate, is critical for sustaining the superconductivity of Nb$_3$Sn and ensuring a low surface resistance under extreme conditions, \textit {e.g.}, high fields and high operating temperatures. We evaluate the Nb/Sn atomic ratios as a function of film depth using XPS combined with ion etching. Additionally, we map the Sn compositions on cutout cross-sections using EDS under STEM after FIB polishing. We compare the results of electrochemically synthesized and vapor-diffused samples.

For electrochemically synthesized Nb$_3$Sn, XPS depth profiling reveals a constant 3:1 ratio at the surface within the 600\,nm region, with the first a few nanometers being affected by oxides, as shown in the representative 60\,nm data in Fig.~\ref{SunFig3}a, and up to 2\,$\mu$m data in Fig.~S18. For example, XPS spectra of the film after sputtering away surface oxides (Fig.~\ref{SunFig3}b) confirm a homogeneous Nb$_3$Sn stoichiometry of 75\,at.\% Nb and 25\,at.\% Sn. Surface EDS spectra (Fig.~S19) taken at different spots on the film verify this ratio. 

Moreover, cross-sectional EDS mapping under STEM, \textit {e.g.}, Fig.~\ref{SunFig3}d, provides a more refined spatial resolution than the XPS probe and confirms the homogeneity of Nb$_3$Sn stochiometry at the surface within hundreds of nanometers. The collective quantification of the distribution of Sn composition over the entire cutout regions of multiple specimens (Fig.~\ref{SunFig3}c) shows a main peak at 25\,at.\% reflecting the prevalence of a homogeneous surface with the desired stoichiometry. A side peak at 18\,at.\% exists and corresponds to the Sn deficiency region appearing at the film base (Fig.~\ref{SunFig3}d). Nevertheless, we observed one 10-nm-sized Sn-deficient part at the surface after examining $>$\,100 grains, which could matter. However, our quantifications confidently prove the predominance of stoichiometric surfaces in the electrochemically synthesized samples. These chemical analyses are consistent with the diffraction and $T_\mathrm{c}$ results (Fig.~\ref{SunFig4}), indicating an ideal stoichiometry.  

In contrast, vapor-diffused Nb$_3$Sn, grown on either pre-anodized or non-anodized Nb surfaces, shows non-homogeneity within the film, with regions containing 25\,at.\% Sn, Sn-rich, and Sn-poor areas (Fig.~\ref{SunFig3}a). Despite the presence of a majority of 25\,at.\%-Sn grains, we observed a three-dimensional non-uniformity of Sn composition and even regions with no film, \textit {e.g.}, as mapped by the cross-sectional STEM/EDS in Fig.~\ref{SunFig3}e. This behavior is further confirmed by a prominent 18\,at.\% peak in the probability distribution of Sn concentration in vapor-diffused samples (Fig.~\ref{SunFig3}c). 

Phase and $T_\mathrm{c}$ calculations \cite{SunRef9,SunRef10,SunRef15,SunRef16} have indicated that tin-poor and tin-rich stoichiometries deteriorate the superconducting properties. For example, Sn concentrations below 20\,at.\% or above 30\,at.\% result in $T_\mathrm{c}$ values below 9\,K (plain Nb). Vapor-diffused Nb$_3$Sn commonly exhibits tin-poor problems owing to (1) an insufficient Sn supply, causing alloying nucleation to follow the 18\,at.\%-Sn boundary (see phase diagram), and (2) slow Sn diffusion within the 18\,at.\%-Sn grains, making rectification difficult. In contrast, electrochemically synthesized Nb$_3$Sn benefits from a sufficient supply of Sn, which promotes nucleation following the 25\,at.\% Sn boundary. 

\section{Structural characterizations and superconducting properties}

\begin{figure}[t!]
\centering
\includegraphics[width=\linewidth]{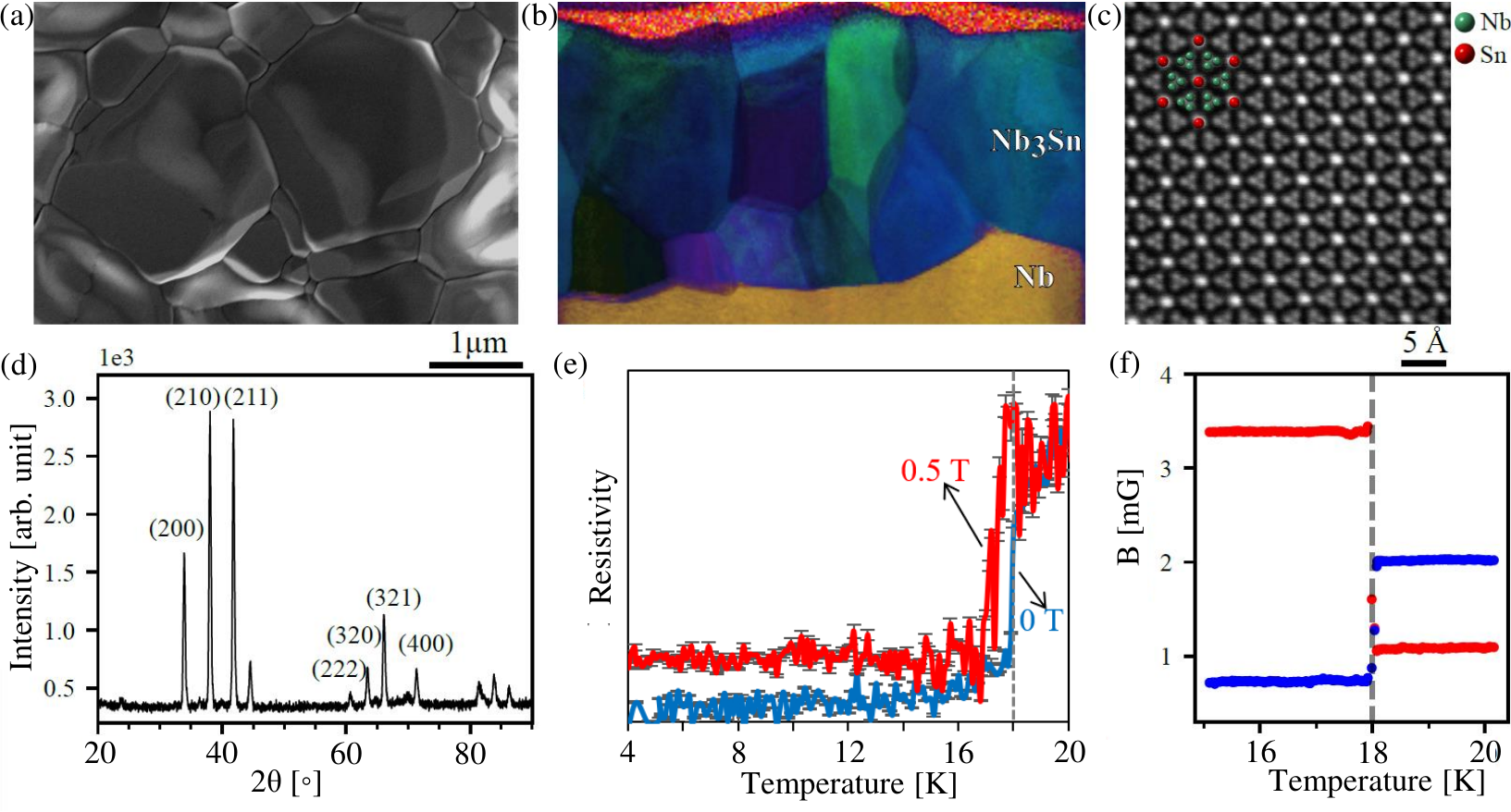}
\caption{\textbf{Physical and structural analysis of Nb$_3$Sn converted from 10\,min electrochemical synthesis.} (a) Surface SEM. (b) Cross-sectional 4D-STEM revealing EWPC-resolved grain contrasts. (c) Atomic resolution STEM. (d) XRD pattern. (e,f) $T_\mathrm{c}$ determined by (e) temperature-dependent resistivity at 0\,T and 0.5\,T fields and (f) flux expulsion measurements. The flux gate magnetometers record an increase or a decrease in the magnetic fields, depending on the orientation of the two magnetometers positioned.}
\label{SunFig4}
\end{figure} 
We further characterize electrochemically synthesized Nb$_3$Sn to study the surface morphology, film thickness, structural phase, grain size and orientation, and superconducting properties (Fig.~\ref{SunFig4}). 

Surface and cross-sectional images, taken by SEM and STEM, show a uniform, smooth 2.4\,$\mu$m-thick Nb$_3$Sn film converted from the 10\,min platings (Fig.~\ref{SunFig4}a,b). Regardless of pre-deposit thicknesses, the surface morphology of converted Nb$_3$Sn barely changes (Fig.~S17) with grain sizes ranging from 200\,nm to 1\,$\mu$m, which are smaller than typical vapor-diffused grains \cite{SunRef38}. Small grains facilitate coordination between orientations and reduce surface roughness. 

Electron diffractions under 4D-STEM, combined with the EWPC transform\cite{SunRef47}, accurately determine the Nb$_3$Sn grains marked by false colors in Fig.~\ref{SunFig4}b. Unlike the predominance of columnar grains extending to the film-substrate interface \cite{SunRef38}, electrochemically synthesized Nb$_3$Sn accommodates multiple grains in the film depth direction. This behavior implies that nucleations, in the presence of pre-deposits wetting on the Nb surface during annealing, likely originate at multiple sites, and the growth develops without being directed to a specific growth direction. In contrast, the heterogeneous nucleation during vapor diffusion strongly relies on the Nb surface conditions, the Sn-vapor adsorption preference, and the Sn diffusion limitations, generating vertically abnormal grains \cite{SunRef9,SunRef73,SunRef48,SunRef38}. This difference in nucleation mechanism may deserve further \textit{in situ} investigations.

STEM atomic images (\textit {e.g.}, Fig.~\ref{SunFig4}c taken from the [111] zone axis) and XRD patterns indexed \cite{SunRef49} (Fig.~\ref{SunFig4}d) confirm an A15 structure following the Pm$\overline{3}$n space group for electrochemically synthesized Nb$_3$Sn. Besides surface oxides, we did not observe foreign phases. 

We determine a $T_\mathrm{c}$ of 18\,K by the resistivity drop (Fig.~\ref{SunFig4}e) and flux expulsion (Fig.~\ref{SunFig4}f) measurements, matching the limit for stoichiometric Nb$_3$Sn. At increasing DC fields, $T_\mathrm{c}$ is sustained at 17\,K at 0.5\,T. Nb$_3$Sn bears critical superheating at 0.4\,T under RF fields, so the slight $T_\mathrm{c}$ degradation at 0.5\,T under DC fields is a positive indicator. 

\section{Outlook: performance demonstration on SRF cavities}

\begin{figure}[t!]
\centering
\includegraphics[width=\linewidth]{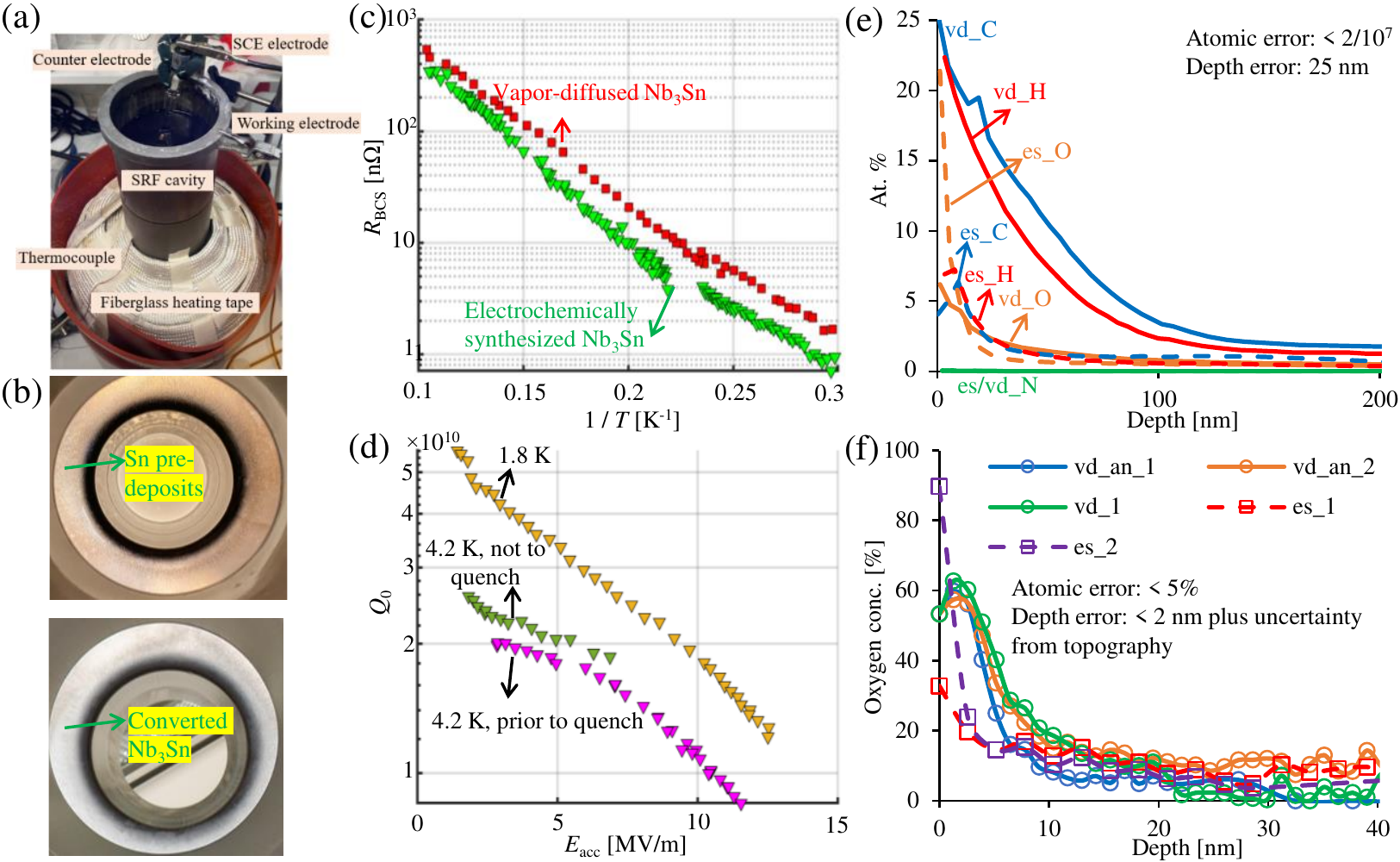}
\caption{\textbf{RF performance of the electrochemically synthesized Nb$_3$Sn 1.3\,GHz SRF cavity, and quantitative analysis of impurities.} (a) Electrochemical deposition setup for the 1.3\,GHz SRF cavity. (b) Pictures of the Sn pre-deposits and converted Nb$_3$Sn on the inner surface of the cavity. The displayed side appears well-deposited, while the other sides show some small areas with a different color \cite{SunRef112}. (c) $R_\mathrm{BCS}$ (total $R_\mathrm{s}$ minus the average of $R_\mathrm{s}$'s at 2.4\,--\,2.6\,K) measured at 2\,MV/m as a function of temperature (1/$\mathit{T}$) for electrochemically synthesized versus vapor-diffused Nb$_3$Sn. (d) $Q_\mathrm{0}$ versus $E_\mathrm{acc}$ at 1.8\,K and 4.2\,K temperatures. The magenta 4.2\,K data were measured prior to quenching, while the 1.8\,K data were measured until quenching occurred. The green 4.2\,K data were measured under a different cooling dynamic, and the cavity remained unquenched. Measurement uncertainties are 10\% in the field and 10\% in the quality factor. (e) SIMS depth profiling of H, C, O, and N, and (f) XPS depth profiling of O, for Nb$_3$Sn made by electrochemical synthesis (es) versus vapor diffusion (vd) with/without pre-anodization (an).}
\label{SunFig6}
\end{figure}

We have successfully scaled up the electrochemical process tailored to a large-scale ($\sim$\,0.2\,m$^2$ surface area), intricate-structured 1.3\,GHz SRF cavity, as shown in Fig.~\ref{SunFig6}a. Fig.~\ref{SunFig6}b demonstrates smooth pre-deposits and a resulting shiny Nb$_3$Sn surface observed on one side of the cavity. During the examination, some small areas on the other side of the cavity displayed a different color \cite{SunRef112}. This suggests the need for further engineering considerations when operating large-volume plating solutions over an extended period. Flipping over the cavity during the deposition process may be a possible solution. Flux expulsion tests on two locations of the cavity (Fig.~\ref{SunFig4}f) confirm the 18\,K $T_\mathrm{c}$ of converted Nb$_3$Sn, suggesting 3:1 stoichiometry. 

Proof-of-concept results demonstrate $R_\mathrm{BCS}$ minimization in the electrochemically synthesized cavity at multiple operating temperatures, approximately two times lower than vapor-diffused cavities (Fig.~\ref{SunFig6}c and Table~\ref{SunTable}). $R_\mathrm{BCS}$ reductions at different temperature regimes reveal the underlying meanings of their corresponding material improvements. At high temperatures ($\textit{e.g.}$, 10\,K), the reduction is linked to the purification in "good" Nb$_3$Sn (Fig.~\ref{SunFig6}e), improving the mean free path. At 4\,K, the lower $R_\mathrm{BCS}$ is strong evidence of the stoichiometric conformity for the absence of Sn-deficient "bad" Nb$_3$Sn, consistent with our chemical analysis. Below 4\,K, the ultra-low $R_\mathrm{BCS}$'s below 1\,n$\Omega$ reflect consistent reductions extrapolated from 4\,K and 10\,K results. 

Benefiting from the low $R_\mathrm{BCS}$, the quality factors at low RF fields are as high as 5.6\,$\times$\,10$^{10}$ measured at 1.8\,K and 2.6\,$\times$\,10$^{10}$ at 4.2\,K (Fig.~\ref{SunFig6}d). Note that the cooldown dynamics can influence the $Q_\mathrm{0}$ for Nb$_3$Sn cavities. The slight difference in 4.2\,K quality factors shown in green and magenta are attributed to small variations in the spatial thermal gradients during cavity cool-down. Nb$_3$Sn cavities require small thermal gradients during cool-down to minimize thermoelectric currents in the bi-metal (Nb$_3$Sn on Nb) cavity wall, which can lead to trapped flux and additional dissipation in RF fields, resulting in lower $Q_\mathrm{0}$ values. The yellow data points represent measurements taken at a helium bath temperature of 1.8\,K, up to quench (maximum field), following the 4.2\,K test shown in green. The quench fields occur at $\sim$\,13\,MV/m, falling within the typical quenching range for vapor-diffused cavities\cite{SunRef75,SunRef79,SunRef18,SunRef39,SunRef31}. Further studies are needed to understand the source(s) of quench mechanisms in electrochemically synthesized Nb$_3$Sn and fully realize the benefits of reduced surface roughness. The observed small spots with a different color, likely indicating roughness, may serve as potential candidates for causing quenching in this cavity at fields similar to those achieved in vapor-diffused Nb$_3$Sn cavities. This hypothesis requires further validation due to the difficulty in characterizing films grown on the inner surface of SRF cavities. Collaborative efforts by SRF labs worldwide to test more electrochemically synthesized cavities would be the most effective way to reveal the actual quench mechanisms. Other hypotheses involve Nb$_3$Sn's small coherence length ($\sim$\,3\,nm), making it sensitive to defects like grain boundaries \cite{SunRef116}, as well as the impact of surface oxides \cite{SunRef114,SunRef104,SunRef117,SunRef118}. We are exploring alternative materials like ZrNb(CO) with a suitable coherence length and low-dielectric-loss ZrO$_2$ \cite{SunRef111}.

To understand the refinement of Nb$_3$Sn achieved, we probed impurities by SIMS (Fig.~\ref{SunFig6}e and Fig.~S20) and XPS (Fig.~\ref{SunFig6}f) depth profiling. H, C, and O concentrations remain essentially low in electrochemically synthesized Nb$_3$Sn, while oxides exist at the surface. The trivial amount of monolayer H and C might be induced by residue from cleaning protocols using methanol or contamination in the air. In contrast, detrimental H and C prevail in vapor-diffused films with high concentrations at considerable depths, significantly affecting the RF penetration region. XPS oxygen profiles show thinner oxides on the electrochemically synthesized Nb$_3$Sn in contrast to vapor-diffused films. Further deconvolution of oxide structures is detailed elsewhere\cite{SunRef104}.

\section{Conclusion}

In summary, we have developed methods to form Nb$_3$Sn for high-performance SRF cavities. We have demonstrated high-quality Nb$_3$Sn with (i) $R_\mathrm{a}$ below 60\,nm on industrial-standard Nb ($\sim$\,100\,nm $R_\mathrm{a}$) which is 5\,$\times$ better than conventional vapor-diffused Nb$_3$Sn; (ii) optimized stoichiometry at the surface 600\,nm region that is larger than the RF field penetration depth (100\,nm); (iii) $T_\mathrm{c}$ and phase confirmed to be stochiometric Nb$_3$Sn; (iv) ultra-low H, C, O, and N impurity concentrations; (v) BCS resistance reduction by approximately 2\,$\times$ at multiple operating temperatures at low fields as compared to vapor-diffused cavities; and (vi) high quality factors at multiple operating temperatures at low fields. The advancements in chemical and surface properties of Nb$_3$Sn are translated to minimizing BCS surface resistances and enhancing quality factors in the SRF cavity. These developments will enable workbench-scale sources of high-energy electrons and X-rays, and will also benefit large particle accelerators to significantly reduce size and cost.

\subsection*{Data availability}
All data generated or analyzed during this study are included in this manuscript and Supplementary Information.

\bibliography{ref}

\noindent\textbf{Acknowledgements}\\
This work was supported by the U.S. National Science Foundation under Award PHY-1549132, the Center for Bright Beams. This work is supported in part by U.S. DOE award DE-SC0008431. This work made use of the Cornell Center for Materials Research Shared Facilities which are supported through the NSF MRSEC program (DMR-1719875) and was performed in part at the Cornell NanoScale Facility, an NNCI member supported by NSF Grant NNCI-2025233. The authors thank Dr. N. Sitaraman, A. C. Hire, Prof. R. Hennig, Prof. T. Arias, and Prof. J. Sethna for valuable discussions through the Center for Bright Beams collaboration. Z.S. acknowledges Dr. K. Dobson for valuable advice on electroplating; T. M. Gruber, H. G. Conklin, P. D. Bishop, Dr. M. Ge, A. Holic, J. Sears, G. Kulina for helping with sample preparation and electrochemical system installation; M. Salim for assisting XPS measurements; and M. Thomas for FIB advice. \\

\noindent\textbf{Author contributions}\\
Z.S. independently developed the methodology; conducted experiments on material growth, processing, optimization, characterization, and superconductivity measurements on sample-scale and cavity studies; and wrote the manuscript. Z.B. performed STEM/EDS experiments and analyses on cross-sectional phase/grain/composition imaging. R.D.P. assisted with the thermal annealing of samples and FFT analysis. L.S. assisted with the annealing and measurement of one cavity. Y.T.S. assisted with STEM imaging. T.O. assisted with RF data analysis. M.O.T., D.A.M., and M.U.L. provided valuable advice to design experiments and understand results mainly through collaboration within the Center for Bright Beams. Z.B., M.O.T., and M.U.L. further revised the manuscript. M.U.L. and D.A.M. acquired funding for the work and supervised this work.\\

\noindent\textbf{Competing interests}\\
The authors have no known competing financial interests.\\

\noindent\textbf{Supplementary information}\\
The Supplementary Information has been submitted as a separate file, including discussions on material challenges in conventional vapor-diffused Nb$_3$Sn, detailed optimization of electrochemical synthesis (surfactant- and precursor-only control studies; CV analyses on the effects of bath temperature, pH, surfactant/precursor ratios, redox potentials, substrate oxide thicknesses, and stirring conditions), SEM and EDS characterization of Sn pre-deposits, and complete datasets of AFM, SEM, XPS, EDS, and SIMS data of electrochemically synthesized Nb$_3$Sn.\\

\end{document}